\documentclass[conference]{IEEEtran}

\usepackage{amsmath, amssymb, graphicx, dsfont, setspace,url, amsfonts, amsthm, multirow}

\newtheoremstyle{Amin}% name
  {3pt}%      Space above
  {3pt}%      Space below
  {}%         Body font
  {}%         Indent amount (empty = no indent, \parindent = para indent)
  {\bfseries}% Thm head font
  {:}%        Punctuation after thm head
  {.5em}%     Space after thm head: " " = normal interword space;
  {}%         Thm head spec (can be left empty, meaning `normal')

\theoremstyle{Amin}
\newtheorem{lemma}{Lemma}
\newtheorem{remark}{Remark}
\newtheorem{defin}{Definition}
\newtheorem{prop}{Proposition}

\newtheorem{theorem}{Theorem}

\usepackage[T1]{fontenc}
\usepackage[latin9]{inputenc}

\usepackage{setspace}

\begin{document}
\setcounter{page}{1}
\title{Symmetric Group Testing and Superimposed Codes}

\author{\IEEEauthorblockN{Amin Emad, Jun Shen and  Olgica Milenkovic}
\authorblockA{University of Illinois, Urbana-Champaign, IL}}
%E-mail: \{emad2,milenkov\}@illinois.edu}
%}
\maketitle

\begin{abstract} We describe a generalization of the group testing problem termed \emph{symmetric group testing}. Unlike in classical binary group testing, the roles played by the input symbols zero and one are ``symmetric'' while the outputs are drawn from a  ternary alphabet. Using an information-theoretic approach, we derive sufficient and necessary conditions for the number of tests required for noise-free and noisy reconstructions. Furthermore, we extend the notion of disjunct (zero-false-drop) and separable (uniquely decipherable) codes to the case of symmetric group testing. For the new family of codes, we derive bounds on their size based on probabilistic methods, and provide construction methods based on coding theoretic ideas. 
\end{abstract}

%%%%%%%%%%%%%%%%%%%%%%%%%%%%%%%%%%%%%%%%%%%%%%%%%%%%%%%%%%%%%%%%%%%%%%%%%%%%%%%%%%%%%%%%%%%%%%%%%%%%%%%%%%%%%%%%%%%%%%%%%%%%%%%%
\vspace{-0.12cm}
\section {Introduction}

Group testing is a combinatorial scheme developed for the purpose of efficient identification of infected individuals in a given pool of subjects~\cite{D43}. The main idea behind the approach is that if a small number of individuals are infected, one can test the population in groups, rather than individually, thereby saving in terms of the number of tests conducted. A tested group is said to be positive if at least one of its members tests positive; otherwise, the tested group is said to be negative. Each individual from a population of size $N$ is represented by a binary test vector (or signature) of length $n$, indicating in which of the $n$ tests the individual participates. The test outcomes are represented by a vector of size $n$ that equals the entry-wise Boolean OR function of the signatures of infected individuals. The reconstruction task consists of identifying the infected individuals based on their test signatures, using the smallest signature length $n$. 

The work in~\cite{D43} was extended in~\cite{KS64}, where the authors proposed two coding schemes for use in information retrieval systems and for channel assignments aimed at relieving congestion in crowded communications bands. The coding schemes are now known as superimposed codes, including \emph{disjunct/zero-false-drop} (D/ZFD) and \emph{separable/uniquely decipherable} (S/UD) codes. For compactness, we henceforth only use the terms disjunct (D) and separable (S) to describe such codes. These two classes of superimposed codes were extensively studied -- see~\cite{DBV02}-\cite{M96} and references therein.

Superimposed codes are inherently asymmetric with respect to the elements of the input alphabet: if all tested subjects are negative (i.e., zero), the output is negative. Otherwise, for any positive number of infected test subjects, the output is positive. Consequently, a zero output carries significantly more information than an output equal to one. For many applications, including DNA pooling and other genomic and biological testing methods with low sensitivity, a test may be positive only if a sufficiently large number of subjects test positive~\cite{CD07}. In particular, a test outcome may be positive if and only if all subjects are infected. 

A probabilistic scheme that makes the group test symmetric with respect to the all-positive and all-negative tests was first described in~\cite{BKS71,H84}. To the best of our knowledge, these are the only two papers dealing with ''symmetric'' group tests, although only within a probabilistic setting where the inputs are assumed to follow a binomial distribution. In addition, the method was only studied for two extremal parameter choices, using a game-theoretic framework in which the player's strategies (i.e., reconstruction methods) are fixed and involve some form of oracle information. 

A more recent approach to the problem of a nearly-symmetric  group testing was described in~\cite{D06}. Threshold group testing assumes that a test is positive (or negative) if more (or less) than $u$ (or $l$) individuals tested are positive, where $1 \leq l \leq u \leq N$. A test produces an \emph{arbitrary} outcome (zero or one) otherwise. The latter feature makes the testing problem highly nontrivial and substantially different from symmetric group testing (SGT).

We are concerned with describing symmetric group testing in a combinatorial setting, extending the concept of symmetry using information theoretic methods and in developing analogues of $D$ and $S$ codes, termed $D_s$ and $S_s$ codes. Our results include bounds on the size of the test set of symmetric group testing, construction methods for symmetric $D$ and $S$ codes, and efficient reconstruction algorithms in the absence and presence of errors. Bounds on the size of $D_s$ and $S_s$ codes are based on Lov\'{a}sz Local Lemma~\cite{AS98} and constructive coding theoretic arguments.

The paper is organized as follows. Section~\ref{sec:info-theory} contains a short exposition regarding symmetric group testing, while information-theoretic bounds on the number of required tests are derived in noisy and noise-free scenarios. In Section~\ref{sec:GGT}, a generalization of SGT (termed \emph{generalized group testing}) is introduced by employing a lower and upper threshold, and information theoretic bounds on the number of required tests are derived. Section~\ref{sec:superimposed} introduces symmetric superimposed codes and contains derivations on the bounds on the size of $D_s$ and $S_s$ codes. Finally, Section~\ref{sec:construction} presents some techniques for constructing symmetric superimposed codes.

%%%%%%%%%%%%%%%%%%%%%%%%%%%%%%%%%%%%%%%%%%%%%%%%%%%%%%%%%%%%%%%%%%%%%%%%%%%%%%%%%%%%%%%%%%%%%%%%%%%%%%%%%%%%%%%%%%%%%%%%%%%%%%%
\vspace{-0.12cm}
\section{Symmetric Group Testing} \label{sec:info-theory}

The use of SGT was originally motivated by applications in circuit testing and chemical component analysis~\cite{BKS71}. As an illustrative example, consider the situation where one is to test $N$ identically designed circuits using only serial and parallel component concatenation. In the serial testing mode, one can detect if all circuits are operational. In the parallel mode, one can detect if all circuits are non-operational. If at least one
circuit is operational and one is non-operational, neither of the two concatenation schemes will be operational. Detecting efficiently which of the circuits are non-operational would require a ternary output group testing scheme.

The reasons for  introducing symmetric group testing with ternary outputs are twofold.  The first motivation is to provide symmetry in the information content of the output symbols zero and one. Note that in standard group testing, a zero output automatically eliminates all tested subjects from further consideration, which is not the case with the symbol one. In symmetric group testing, the symbols zero and one play a symmetric role. The second motivation comes from biological applications in which the sensitivity of the measurement devices is such that it can only provide a range for the number of infected individuals in a pool: for example, the output may be ``0'' if less than $l$ individuals in the test set are infected, ''1'' if more than $u \geq l$ individuals are infected, and ``2'' in all other cases.

Throughout the paper we use the word ``positive'' (``negative'') to indicate that a tested subject has (does not have) a given property. For the asymmetric testing strategy, the outcome of a test is negative if all tested subjects are negative, and positive otherwise. The outcome of a symmetric test is said to be positive (denoted by ``1''), inconclusive (denoted by ``2''), or negative (denoted by ``0''), if all the subjects tested are positive, at least one subject is positive and another one is negative, and all subjects are negative, respectively. 

Let $N$ denote the total number of test subjects, and let $\mathcal{D}$ denote the defective set with cardinality $|\mathcal{D}|=m$. Furthermore, let $\textbf{X}_{\mathcal{D}}$ denote the collection of codewords (or signatures) corresponding to $\mathcal{D}$; note that the length of each codeword is equal to $n$. Also, let $\textbf{y}\in\{0,1,2\}^n$ denote the noise-free observation vector (test outcome) equal to the \emph{ternary addition} of the codewords of the defective set, where ternary addition is defined as follows. \begin{defin}\label{DEF1} For a ternary alphabet $\{{0,1,2\}}$, we define \emph{ternary addition}, $+$, via the rules $0+0=0$, $1+1=1$, and $0+1=1+0=0+2=2+0=1+2=2+1=2+2=2$. Clearly, ternary addition is commutative and associative.
\end{defin}

Note that in general the ternary addition operator is more informative than its binary counterpart; consequently, one expects that the number of required tests in the SGT is smaller than the number of required tests in a similar asymmetric group testing (AGT) scheme. We hence focus on finding upper and lower bounds on the minimum number of tests in a SGT scheme that ensures detection of the defective set with probability of error asymptotically converging to zero. 

%%%%%%%%%%%%%%%%%%%%%%%%%%%%%%%%%%%%%%%%%%%%%%%%%%%%%%%%%%%%%%%%%%%%%%%%%%%%%%%%%%%%%%%%%%%%%%%%%%
\vspace{-0.12cm}
\subsection{Noise-free Symmetric Group Testing}
In the noise-free case, the observation vector is equal to the  superposition of the signatures of defectives, i.e., 
\vspace{-0.1cm}
\begin{equation}
\textbf{y}=\textbf{x}_{i_1}+\textbf{x}_{i_2}+\cdots+\textbf{x}_{i_m}\ \ \ \ \forall i_j\in\mathcal{D},
\vspace{-0.1cm}
\end{equation}
where ``$+$'' denotes ternary addition, and $\textbf{x}_{i_j}$ stands for the signature of the $j^{\textbf{\textnormal{th}}}$ defective subject. 

Let the tests be designed independently, and let $p$ denote the probability that a subject is part of a given test, independent of all other subjects. 
It was shown in \cite{AS10} that for any $m=o(N)$, a sufficient number of tests for asymptotically achieving a probability of error equal to zero in the AGT setting is lower bounded as
\vspace{-0.1cm}
\begin{equation}\label{sufficient}
n>\max_{i:(\mathcal{D}_1,\mathcal{D}_2)\in\mathcal{A}_{\mathcal{D}}^{(i)}}\frac{\log{{N-m}\choose i}{m\choose i}}{I(X_{\mathcal{D}_1};X_{\mathcal{D}_2},y)}\ \ \ \ i=1,2,\cdots,m
\vspace{-0.1cm}
\end{equation}
where $\mathcal{A}_{\mathcal{D}}^{(i)}$ denotes all ordered pairs of sets $(\mathcal{D}_1,\mathcal{D}_2)$ that partition the defective set,  such that $|\mathcal{D}_1|=i$ and $|\mathcal{D}_2|=m-i$. 
In the above equation, $I(X_{\mathcal{D}_1};X_{\mathcal{D}_2},y)$ stands for the mutual information between $X_{\mathcal{D}_1}$ and $(X_{\mathcal{D}_2},y)$; for a single test, $X_{\mathcal{D}_t}$ (where $t=1,2$) is a vector of size $1\times |\mathcal{D}_t|$, with its $k^{\textnormal{th}}$ entry equal to 1 if the $k^{\textnormal{th}}$ defective subject in $\mathcal{D}_t$ is in the test and 0 otherwise, while $y$ is the outcome of the test. 

Also, a necessary condition for zero error probability in AGT was shown to be of the form  \cite{AS10}
\vspace{-0.1cm}
\begin{equation}\label{necessary}
n\geq\max_{i:(\mathcal{D}_1,\mathcal{D}_2)\in\mathcal{A}_{\mathcal{D}}^{(i)}}\frac{\log{{N-m+i}\choose i}}{I(X_{\mathcal{D}_1};X_{\mathcal{D}_2},y)}\ \ \ \ i=1,2,\cdots,m.
\vspace{-0.1cm}
\end{equation}
By following the same steps in the proof of the above two bounds, it can be easily shown that the same sufficient and necessary conditions hold for the SGT scheme, except for the fact that the mutual information will evaluate to a different form given the change in the output alphabet. 
Furthermore, it can be easily shown that for a fixed value of $m$, $\log{{N-m+i}\choose i}\sim i\log{N}$ and $\log{{N-m}\choose i}{m\choose i}\sim i\log{N}$, as $N\rightarrow\infty$; consequently, these bounds are asymptotically tight. 

In the following Proposition, we evaluate the expressions for mutual information in \eqref{sufficient} and \eqref{necessary} for SGT.

\begin{prop}
In the noise-free SGT for $m\geq 2$,
\vspace{-0.1cm}
\begin{align}\label{I_SGT}
&I_{\textnormal{S}}(X_{\mathcal{D}_1};X_{\mathcal{D}_2},y)\\\nonumber
&=\!\left\{
     \begin{array}{lr}
       \!\!\!(1\!-\!p)^{m-i}h\!\left((1\!-\!p)^i\right)\!+\!p^{m-i}h\!\left(p^i\right) & \textnormal{if}\  1\leq i\leq m-1\\
       \!\!\! h\left(p^m,(1-p)^m\right) & \textnormal{if}\  i=m\\
     \end{array}\!,
   \right.
\end{align}
where $|\mathcal{D}_1|=i$, and where for any $0\leq \zeta,\gamma\leq1$, $h(\zeta)=-\zeta\log \zeta-(1-\zeta)\log(1-\zeta)$ and $h(\zeta,\gamma)=-\zeta\log\zeta-\gamma\log\gamma-(1-\zeta-\gamma)\log(1-\zeta-\gamma)$.
\end{prop}
\begin{IEEEproof}
Let $w_j$ (where $j=1,2$) denote the number of ones in $X_{\mathcal{D}_j}$, or alternatively, the Hamming weight of $X_{\mathcal{D}_j}$. 
From the definition of mutual information, one has
\vspace{-0.1cm}
\begin{align}\nonumber
&I_{\textnormal{S}}(X_{\mathcal{D}_1};X_{\mathcal{D}_2},y)=H(y|X_{\mathcal{D}_2})-H(y|X_{\mathcal{D}})=H(y|X_{\mathcal{D}_2})\\\nonumber
&=\!-\!\sum_{j=0}^{2}\!\sum_{k=0}^{m-i}\!P(w_2\!=\!k)P(y\!=\!j|w_2\!=\!k)\log P(y\!=\!j|w_2\!=\!k).
\end{align}
If $1\leq i\leq m-1$, then
\vspace{-0.1cm}
\begin{align}\nonumber
&I_{\textnormal{S}}(X_{\mathcal{D}_1};X_{\mathcal{D}_2},y)\\\nonumber
&=-P(w_2=0)P(y=0|w_2=0)\log P(y=0|w_2=0)\\\nonumber
&-P(w_2=0)P(y=2|w_2=0)\log P(y=2|w_2=0)\\\nonumber
&-P(w_2\!=\!m\!-\!i)P(y\!=\!2|w_2\!=\!m\!-\!i)\log P(y\!=\!2|w_2\!=\!m\!-\!i)\\\nonumber
&-P(w_2\!=\!m\!-\!i)P(y\!=\!1|w_2\!=\!m\!-\!i)\log P(y\!=\!1|w_2\!=\!m\!-\!i)\\\nonumber
&=(1-p)^{m-i}h\left((1-p)^i\right)+p^{m-i}h\left(p^i\right).
\end{align}
Otherwise, if $i=m$, then
\vspace{-0.1cm}
\begin{align}\nonumber
&I_{\textnormal{S}}(X_{\mathcal{D}_1};X_{\mathcal{D}_2},y)=-(1-p)^m\log\left((1-p)^m\right)-p^m\log\left(p^m\right)\\\nonumber
&\  \  \  -\left(1-p^m-(1-p)^m\right)\log\left(1-p^m-(1-p)^m\right).
\end{align}
\end{IEEEproof}
\noindent Similarly, it can be shown that for AGT, 
\vspace{-0.1cm}
\begin{equation}\label{I_AGT}
I_{\textnormal{A}}(X_{\mathcal{D}_1};X_{\mathcal{D}_2},y)=(1-p)^{m-i}h\left((1-p)^i\right).
\vspace{-0.1cm}
\end{equation}
In order to compare SGT and AGT, fix $p$ and let 
\vspace{-0.1cm}
\begin{equation}\label{alpha_p}
\alpha(m,p)\!=\!\!\!\!\!\!\min_{i:(\mathcal{D}_1,\mathcal{D}_2)\in\mathcal{A}_{\mathcal{D}}^{(i)}}\frac{I(X_{\mathcal{D}_1};X_{\mathcal{D}_2},y)}{i}\ \ i\!=\!1,2,\cdots,m.
\vspace{-0.1cm}
\end{equation}
Using the above definition, the bounds in (\ref{sufficient}) and (\ref{necessary}) asymptotically simplify to $n>\frac{\log N}{\alpha(m,p)}$ and $n\geq\frac{\log N}{\alpha(m,p)}$, respectively. Also, let $\alpha_{\textnormal{A}}(m)$ and $\alpha_{\textnormal{S}}(m)$ denote the values of $\alpha(m,p)$ for the choice of $p$ that minimizes the lower bounds in the AGT and SGT expressions\footnote{Note that for a fixed value of $m$, the value of $p$ that maximizes $\alpha(m,p)$ may differ for AGT and SGT.}, respectively. Fig. 1 shows the behavior of these parameters with respect to $m$. As can be seen, for $m=2$, $\frac{\alpha_{\textnormal{S}}(m)}{\alpha_{\textnormal{A}}(m)}=1.5$, but as $m$ grows, this ratio converges to one. 

\begin{figure}\label{figure1}
\begin{center}
\includegraphics[width=7.5cm]{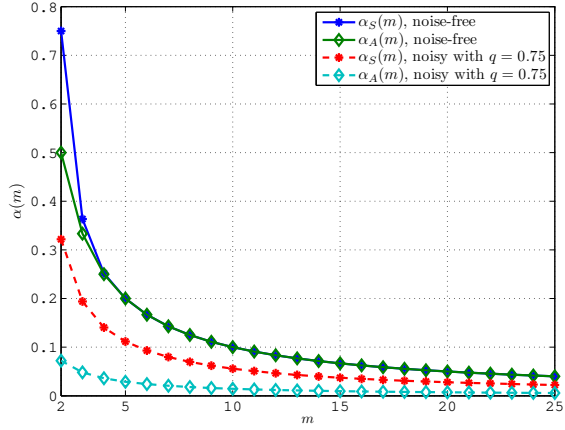}
\end{center}
\vspace*{-15pt}
\caption{The behavior of $\alpha_{\textnormal{S}}(m)$ and $\alpha_{\textnormal{A}}(m)$ versus $m$ in the noise-free and noisy scenario.}
%\vspace*{-23pt}
\vspace*{-15pt}
\end{figure}

%%%%%%%%%%%%%%%%%%%%%%%%%%%%%%%%%%%%%%%
\vspace{-0.12cm}
\subsection{Symmetric Group Testing in the Presence of Binary Additive Noise}
This type of noise accounts for false alarms in the outcome of the tests. In this case, the noise vector, $\textbf{z}$, is modeled as a vector of independent identically distributed (i.i.d.) Bernoulli random variables with parameter $q$; the vector of noisy observations, $\tilde{\textbf{{y}}}$, equals the ternary sum of the vector of noise-free observations and the noise vector, i.e., $\tilde{\textbf{y}}=\textbf{y}+\textbf{z}$. 
Note that in this model, we assume that both 0's and 1's may change to the value 2, while 2's remain unaltered in the presence of noise. 
This model applies whenever dilution effects may occur, since adding one positive (negative) subject may change the outcome of a negative (positive) group.

\begin{prop}
In the presence of binary additive noise, for any $m\geq2$,\vspace{-0.1cm}
\begin{align}\label{I_SBAN}
&I_{\textnormal{S}}(X_{\mathcal{D}_1};X_{\mathcal{D}_2},\tilde{y})\\\nonumber
&=\!\!\left\{
     \begin{array}{lr}
      \!\!\!\! (1-p)^{m-i}h\!\left((1-p)^i(1-q)\right)+p^{m-i}h\!\left(p^iq\right)\\
\  \  -\left[(1-p)^m+p^m\right]h(q) \  \  \  \  \  \  \  \  \  \  \  \  \textnormal{if}\  \  \  \  1\!\leq\! i\!\leq\! m\!-\!1\\
 	          \\
       \!\!\!\! h\left(p^mq,(1-p)^m(1-q)\right)\\
\  \   -\left[(1-p)^m+p^m\right]h(q) \  \  \  \  \  \  \  \  \  \  \  \  \textnormal{if}\  \  \  \  i\!=\!m\\
     \end{array}\!,
   \right.
\end{align}
\end{prop}
\vspace{-0.1cm}
\begin{IEEEproof}
It can be easily shown that if $1\leq i\leq m-1$,
\vspace{-0.1cm}
\begin{align}\nonumber
&H(\tilde{y}|X_{\mathcal{D}_2})\\\nonumber
&=\!-\!\sum_{j=0}^{2}\sum_{k=0}^{m-i}P(w_2\!=\!k)P(\tilde{y}\!=\!j|w_2\!=\!k)\log P(\tilde{y}\!=\!j|w_2\!=\!k)\\\nonumber
&=(1-p)^{m-i}h\left((1-p)^i(1-q)\right)+p^{m-i}h\left(p^iq\right),
\end{align}
where $w_2$ denotes the Hamming weight of $X_{\mathcal{D}_2}$.
Similarly, if $i=m$, then $H(\tilde{y}|X_{\mathcal{D}_2})=h\left(p^mq,(1-p)^m(1-q)\right)$. Also one has
\vspace{-0.1cm}
\begin{align}\nonumber
H(\!\tilde{y}|X_{\mathcal{D}}\!)&\!=\!-\!\!\sum_{j=0}^{2}\!\sum_{k=0}^{m-i}\!P(\!w\!=\!k\!)P(\!\tilde{y}\!=\!j|w\!=\!k\!)\log\! P(\!\tilde{y}\!=\!j|w\!=\!k\!)\\\nonumber
&=\left[(1-p)^m+p^m\right]h(q),
\end{align}
where $w$ denotes the Hamming weight of $X_{\mathcal{D}}$.
Using these two expressions in the definition of the mutual information completes the proof.
\end{IEEEproof}
Similarly, it can be shown that for AGT,
\vspace{-0.1cm}
\begin{align}\label{I_ABAN}
I_{\textnormal{A}}(X_{\mathcal{D}_1};X_{\mathcal{D}_2},\tilde{y})=&(1-p)^{m-i}h\!\left((1-p)^i(1-q)\right)\\\nonumber
&-(1-p)^mh(q).
\end{align}
Fig. 1 also shows the behavior of $\alpha_{\textnormal{S}}(m)$ and $\alpha_{\textnormal{A}}(m)$ with respect to $m$, for the dilution noise model with $q=0.75$. As can be seen, SGT outperforms the AGT scheme: for $m=2$, the ratio $\frac{\alpha_{\textnormal{S}}(m)}{\alpha_{\textnormal{A}}(m)}$ is approximately equal to $4.5$, and as $m$ grows it decreases. Note that for small values of $q$ in this model, SGT does not offer any significant advantage when compared to AGT: for $q=0$, eqs. (\ref{I_SBAN}) and (\ref{I_ABAN}) are identical.

%%%%%%%%%%%%%%%%%%%%%%%%%%%%%%%%%%%%%%%%%%%%%%%%%%%%%%%%%%%%%%%%%%%%%%%%%%%%%%%%%%%%%%%%%%%%%%%%%%%%%%%%%%%%%%%%%%%%%%%%%%%%%%%%%%%%%%%%
\vspace{-0.12cm}
\section{Generalized Group Testing}\label{sec:GGT}
In many applications, the experiment cannot be modeled by AGT or SGT, and a more general model is required. We hence consider the generalized group testing (GGT) problem in which the outcome of a test equals 0 if the number of defectives in the corresponding pool is less than or equal to $\eta_1$ (where $\eta_1\geq0$), equals 1 if the number of defectives is larger than $\eta_2$ (where $\eta_1\leq\eta_2\leq m-1$), and equals 2 if the number of defectives is larger than $\eta_1$ and is less than or equal to $\eta_2$. Note that when $\eta_1=\eta_2=0$ GGT reduces to AGT, and when $\eta_1=0$ and $\eta_2=m-1$ GGT reduces to SGT.

\begin{prop}
In GGT, and for any $i=1,2,\cdots,m$, where $m\geq2$,
\begin{subequations}
\vspace{-0.1cm}
\begin{align}\nonumber
I_{\textnormal{G}}(X_{\mathcal{D}_1};X_{\mathcal{D}_2},{y})=&\!\!\sum_{k=0}^{\min(\eta_1,m-i)}\!\!p_0(k)\  h\left(p_1(k,\eta_1),p_2(k,\eta_2)\right)\\
&+\sum_{k=\eta_1+1}^{\min(\eta_2,m-i)}p_0(k)\ h\left(p_2(k,\eta_2)\right)
\end{align}
with
\vspace{-0.1cm}
\begin{equation}
p_0(k)={m-i\choose k}p^k(1-p)^{m-i-k},
\vspace{-0.1cm}
\end{equation}
\vspace{-0.1cm}
\begin{equation}
p_1(k,\eta_1)=\sum_{l=0}^{\min(i,\eta_1-k)}{i\choose l}p^l(1-p)^{i-l},
\vspace{-0.1cm}
\end{equation}
\vspace{-0.1cm}
\begin{equation}
p_2(k,\eta_2)=1-\sum_{l=0}^{\min(i,\eta_2-k)}{i\choose l}p^l(1-p)^{i-l}.
\vspace{-0.1cm}
\end{equation}
\end{subequations}
As before, $p$ denotes the probability that a subject is part of a given random test. 
\end{prop}

Similarly to the case of AGT and SGT, one can define $\alpha_{\textnormal{G}}(m)$ as
\vspace{-0.1cm}
\begin{equation}
\alpha_{\textnormal{G}}(m)=\max_{p,\eta_1,\eta_2}\alpha(m,p,\eta_1,\eta_2)
\vspace{-0.1cm}
\end{equation}
where $\alpha(m,p,\eta_1,\eta_2)$ is defined in the same manner as $\alpha(m,p)$ in (\ref{alpha_p}). Fig. 2 shows $\alpha(m)$ versus $m$ for AGT, SGT, and GGT. As can be seen, when $m\geq3$, GGT outperforms SGT and AGT. For example, when $m=3$, one has $\frac{\alpha_{\textnormal{G}}(m)}{\alpha_{\textnormal{S}}(m)}\approx1.4$ 
and $\frac{\alpha_{\textnormal{G}}(m)}{\alpha_{\textnormal{A}}(m)}\approx1.6$, and at $m=25$, one has $\frac{\alpha_{\textnormal{G}}(m)}{\alpha_{\textnormal{S}}(m)}\approx\frac{\alpha_{\textnormal{G}}(m)}{\alpha_{\textnormal{A}}(m)}\approx1.6$. The arguments that maximize $\alpha(m,p,\eta_1,\eta_2)$ are denoted by $p^*$, $\eta_1^*$, and $\eta_2^*$ and are tabulated in Table 1 for different values of $m$. 

\begin{figure}
\begin{center}
\includegraphics[width=7.5cm]{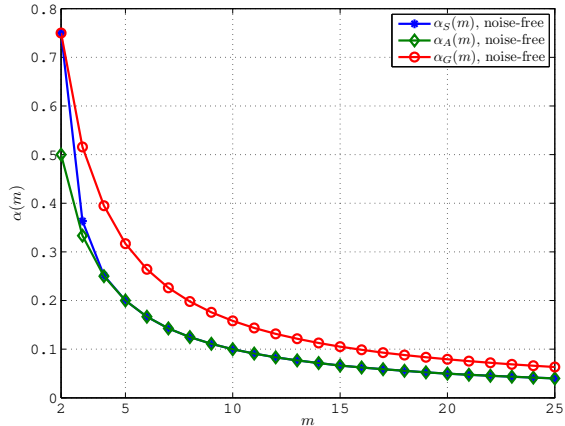}
\end{center}
\vspace*{-15pt}
\caption{The behavior of $\alpha_{\textnormal{G}}(m)$, $\alpha_{\textnormal{S}}(m)$, and $\alpha_{\textnormal{A}}(m)$ versus $m$ in the absence of noise.}
%\vspace*{-23pt}
\vspace*{-15pt}
\end{figure}

\begin{table}[t!]
	\centering
	\caption{Optimal GGT probabilities and thresholds for different values of $m$.}
		\begin{tabular}{|c|c|c|c||c|c|c|c|}
			\hline 
			$m$ &  $p^*$ & $\eta_1^*$ & $\eta_2^*$	& $m$ &  $p^*$ & $\eta_1^*$ & $\eta_2^*$\\ 
			\hline\hline
			
			2 & 0.500 & 0 & 1 & \multirow{2}{*}{12} & 0.175 & 1 & 2\\	\cline{1-4}	\cline{6-8}			
				
			\multirow{2}{*}{3} & 0.351 & 0 & 1 & 	&	0.825	&	9	&	10 \\	\cline{2-8}
			
				&	0.649	&	1	&	2	& \multirow{2}{*}{13} & 0.161 & 1 & 2\\	\cline{1-4} \cline{6-8}

			4	&	0.500	&	1	&	2	& &	0.839	&	10	&	11\\	\hline
			
			\multirow{2}{*}{5} & 0.406 & 1 & 2 & \multirow{2}{*}{14} & 0.150 & 1 & 2\\	\cline{2-4}\cline{6-8}
				&	0.594	&	2	&	3 & &	0.850 &	11	&	12\\	\hline
			 
			\multirow{2}{*}{6} & 0.341 & 1 & 2 & \multirow{2}{*}{15} & 0.076 & 0 & 1\\	\cline{2-4}\cline{6-8}
				&	0.659	&	3	&	4 & &	0.924	&	13	&	14 \\	\hline
				
			\multirow{2}{*}{7} & 0.294 & 1 & 2 & \multirow{2}{*}{16} &	0.071  & 0 & 1\\	\cline{2-4}\cline{6-8}
				&	0.706	&	4	&	5 & &	0.929	&	14	&	15\\	\hline
				
			\multirow{2}{*}{8} & 0.259 & 1 & 2 & 17 & 0.500 & 7 & 9\\	\cline{2-8}
			
				&	0.741	&	5	&	6 & \multirow{2}{*}{18} & 0.473 & 7 & 9\\	\cline{1-4}\cline{6-8}

			\multirow{2}{*}{9} & 0.231 & 1 & 2 & &	0.527	&	8	&	10\\	\cline{2-8}
				&	0.769	&	6	&	7 &	19 & 0.500 & 8 & 10 \\	\hline				
				
			\multirow{2}{*}{10} & 0.209 & 1 & 2 & \multirow{2}{*}{20} & 0.475 & 8 & 10\\	\cline{2-4}\cline{6-8}
				&	0.791	&	7	&	8 & &	0.525	&	9	&	11\\	\hline

			\multirow{2}{*}{11} & 0.190 & 1 & 2 & 21 & 0.500 & 9 & 11\\	\cline{2-8}
				&	0.810	&	8	&	9 \\	\cline{1-4}				
														
			\end{tabular}
			\vspace*{-15pt}	
\end{table}

%%%%%%%%%%%%%%%%%%%%%%%%%%%%%%%%%%%%%%%%%%%%%%%%%%%%%%%%%%%%%%%%%%%%%%%%%%%%%%%%%%%%%%%%%%%%%%%%%%%%%%%%%%%%%%%%%%%%%%%%%%%%%%%%%%%%%%%%%%%%%%%%%%%
\vspace{-0.12cm}	
\section{Symmetric Superimposed Codes}\label{sec:superimposed}
In this section, we introduce disjunct and separable symmetric superimposed codes for the SGT scheme. We postpone the discussion of generalized disjunct and separable codes to the full version of the paper.

\begin{defin} Given two ternary vectors $\textbf{x}$ and $\textbf{y}$, and a ternary addition operator, we say that $\textbf{y}$ is \emph{included} in $\textbf{x}$ if $\textbf{x}+\textbf{y}=\textbf{x}$.
\end{defin}

\begin{defin} Let the addition operator be the ternary addition described in Def. \ref{DEF1}. 
A code $\mathcal{C}$ is a symmetric $m$-disjunct code if for any sets of binary codewords $\{\textbf{x}_{1},\textbf{x}_{2},\cdots,\textbf{x}_{s}\}$ and $\{\textbf{y}_{1},\textbf{y}_{2},\cdots,\textbf{y}_{t}\}$, the sum $\textbf{x}_{1}+\textbf{x}_{2}+...+\textbf{x}_{s}$ being included in $\textbf{y}_{1}+\textbf{y}_{2}+...+\textbf{y}_{t}$ implies that $\{\textbf{x}_{1},\textbf{x}_{2},\cdots,\textbf{x}_{s}\}\subseteq\{\textbf{y}_{1},\textbf{y}_{2},\cdots,\textbf{y}_{t}\}$, for any $s,t\leq m$. 
\end{defin}

\begin{defin} A symmetric code $\mathcal{C}$ is an $m$-separable code if $\textbf{x}_{1}+\textbf{x}_{2}+\cdots+\textbf{x}_{s}=\textbf{y}_{1}+\textbf{y}_{2}+...+\textbf{y}_{t}$ implies $s=t$ and $\{\textbf{x}_{1},\textbf{x}_{2},\cdots,\textbf{x}_{s}\}=\{\textbf{y}_{1},\textbf{y}_{2},\cdots,\textbf{y}_{t}\}$ for any $s,t \leq m$. 
\end{defin}

Henceforth, we refer to $m$ as the strength of code $\mathcal{C}$ and use $N(m)$ and $n(m)$ to denote the number of codewords (the signatures of the subjects) and their length, respectively. The rate of a superimposed code of strength $m$ is defined as $R_m=\frac{\log N(m)}{n(m)}$, where $\log$ stands for the logarithm base two. Whenever apparent from the context, we use $N$ and $n$ instead of $N(m)$ and $n(m)$.

%%%%%%%%%%%%%%%%%%%%%%%%%%%%%%%%%%%%%%%%%%
\vspace{-0.12cm}
\subsection{Bounds on the Size of Symmetric Disjunct Codes}
In this subsection, we derive an upper bound on the size of $D_s(m)$ codes using probabilistic methods. 

\begin{theorem}[Upper Bound]\label{Thm_upper}
Let $m\geq2$ be a fixed number and let $n\rightarrow\infty$; if $N$ is asymptotically upper bounded by $A{B_{\textnormal{S}}}^n$, where $A=\left[\frac{m!}{(m+1)^{2}e}\right]^{1/m}$ and $B_{\textnormal{S}}=\left[\frac{2^{m}}{2^{m}-1}\right]^{1/m}>1$, then there exists a symmetric disjunct superimposed $(n,N,m)-$code. The rate of this code is asymptotically upper bounded by $R_{\textnormal{S}}(m)=\frac{\log N}{n}\sim \log B_{\textnormal{S}}>0$. 
\end{theorem}

\begin{IEEEproof}
Let $\mathcal{C}$ be a set of $N$ codewords with length $n$, and let $\mathcal{M}$ be a set of $m+1$ codewords chosen from $\mathcal{C}$. There are $\binom{N}{m+1}$ different possibilities for $\mathcal{M}$. For the $i^{\textnormal{th}}$ possible choice of $\mathcal{M}$, define $E_{i}$ as the event that at least one of the codewords in $\mathcal{M}$ is included in the ternary sum of the other $m$ codewords. From this definition, each $E_{i}$ is mutually independent of all except $\left[\binom{N}{m+1}-\binom{N-(m+1)}{m+1}-1\right]$ other events. Suppose that $P(E_i)\leq p'$ for all $i$. Using Lov\'{a}sz's local lemma~\cite{AS98}, if 
\vspace{-0.1cm}
\begin{equation}\label{lovasz}
ep'\left[\binom{N}{m+1}-\binom{N-(m+1)}{m+1}\right]<1,
\vspace{-0.1cm}
\end{equation}
then $P\left(\bigcap_{i}\overline{E}_{i}\right)>0$, where $e$ is the base of the natural logarithm and $\overline{E}_{i}$ is the complement of the event $E_i$. 
In other words, if (\ref{lovasz}) is satisfied, there exists a choice of $N$ codewords that form a symmetric disjunct superimposed $(n,N,m)-$code, $\mathcal{C}$. 
Note that if no codeword is included in the sum of $m$ other codewords, then no codeword is included in the sum of any set of other codewords with cardinality smaller than $m$.

\begin{lemma}\label{upper_p}
For all $i$ one has
\vspace{-0.1cm}
\begin{equation}
P(E_i)\leq p'=1-\frac{2^{n(m+1)}-(m+1)\left(2^{m+1}-2\right)^n}{(m+1)!{2^n\choose m+1}},
\vspace{-0.1cm}
\end{equation}
and as $n\rightarrow\infty$, one has
\vspace{-0.1cm}
\begin{equation}
p'\sim (m+1)  \left[1 - \left(\frac{1}{2}\right)^m \right] ^ n.
\vspace{-0.1cm}
\end{equation}
\end{lemma}

Substituting $p'$ into (\ref{lovasz}) and using the fact that $\binom{N}{m+1}-\binom{N-(m+1)}{m+1}\sim\frac{(m+1)^{2}N^{m}}{(m+1)!}$, as $N\rightarrow\infty$, completes the proof.
\end{IEEEproof}

\begin{remark}
Similarly, it can be proved that for $m\geq2$, if $N$ is asymptotically smaller than $A{B_{\textnormal{A}}}^n$, where $B_{\textnormal{A}}=\left[\frac{2^{m+1}}{2^{m+1}-1}\right]^{1/m}>1$, then there exists an asymmetric disjunct superimposed $(n,N,m)-$code. The rate of this code is asymptotically upper bounded by $R_{\textnormal{A}}(m)=\frac{\log N}{n}\sim \log B_{\textnormal{A}}>0$. Consequently, 
\vspace{-0.1cm}
\begin{equation}
\frac{R_{\textnormal{S}}(m)}{R_{\textnormal{A}}(m)}=\frac{m-\log\left(2^m-1\right)}{m+1-\log\left(2^{m+1}-1\right)}.
\vspace{-0.1cm}
\end{equation}
The ratio $\frac{R_{\textnormal{S}}(m)}{R_{\textnormal{A}}(m)}$ is approximately equal to 2 when $m=2$ and tends to 1 as $m$ grows. 
\end{remark}

%%%%%%%%%%%%%%%%%%%%%%%%%%%%%%%%%%%%%%%%%%%%%%%%%%%%%%%%%%%%%%%%%%%%%%%%%%%%%%%%%%%%%%%%
\vspace{-0.12cm}
\subsection{Bounds on the Size of Symmetric Separable Codes}
In this subsection, we find an upper bound on the size of symmetric separable codes when $m=2$.

\begin{theorem}[Upper Bound]\label{Thm_upper2}
Let $m=2$ and let $n\rightarrow\infty$; if $N$ is asymptotically smaller than $A'{{B'}_{\textnormal{S}}}^n$, where $A'=(2e)^{-\frac{1}{3}}\approx0.569$ and ${B'}_{\textnormal{S}}=\left(\frac{8}{3}\right)^{\frac{1}{3}}\approx1.387$, then there exists a symmetric separable superimposed $(n,N,2)-$code.  
\end{theorem}

\begin{IEEEproof}
No two distinct pairs of codewords of a separable code have an identical ternary sum. Let $\mathcal{C}$ be a code of $N$ codewords with length $n$ and let 
$\mathcal{M}$ be a set of four codewords of $\mathcal{C}$. There are $\binom{N}{4}$ different choices for $\mathcal{M}$. For the $i^{\textnormal{th}}$ choice of 
$\mathcal{M}$, we define $E_i$ as the event that at least two distinct pairs of codewords in $\mathcal{M}$ have an identical ternary sum. Using Lov\'{a}sz's local lemma, if 
\vspace{-0.1cm}
\begin{equation}\label{lovasz2}
ep''\left[\binom{N}{4}-\binom{N-4}{4}\right]<1,
\vspace{-0.1cm}
\end{equation}
then $P\left(\bigcap_{i}\overline{E}_{i}\right)>0$, where $p''$ is an upper bound on $P(E_i)$, i.e. $P(E_i)\leq p''$, for all $i$. 

\begin{lemma}
One has
\vspace{-0.1cm}
\begin{equation}\nonumber
p''=\frac{3\cdot6^{n}-6\cdot4^{n}+3\cdot2^{n}}{2^{n}(2^{n}-1)(2^{n}-2)(2^{n}-3)},
\vspace{-0.1cm}
\end{equation}
and as $n\rightarrow\infty$, $p''\sim3\left(\frac{3}{8}\right)^{n}$. 
\end{lemma}

Substituting $p''$ into (\ref{lovasz2}) and using the fact that $\binom{N}{4}\!-\!\binom{N-4}{4}\sim\frac{{2}}{3}N^{3}$, as $N\rightarrow\infty$, completes the proof.
\end{IEEEproof}

%%%%%%%%%%%%%%%%%%%%%%%%%%%%%%%%%%%%%%%%%%%%%%%%%%
\vspace{-0.12cm}
\section{Construction of Symmetric Superimposed Codes}\label{sec:construction}

Due to space limitations, we only consider symmetric separable codes with $m=2$ for which a construction based on coding theoretic methods is particularly simple.
\begin{prop}
The columns of a parity-check matrix of a linear code, with minimum distance of at least five, form a symmetric $UD_{2}$ code.\end{prop}
\begin{IEEEproof}
Let $\textbf{c}_{1},\textbf{c}_{2},\textbf{c}_{3},\textbf{c}_{4}$ denote four different columns of the parity-check matrix $H.$ 
Furthermore, assume that $\textbf{c}_{1}+\textbf{c}_{2}=\textbf{c}_{3}+\textbf{c}_{4}=\textbf{c}$. At the positions where $\textbf{c}$ is $0$ or
$1,$ all codewords $\textbf{c}_{1},\textbf{c}_{2},\textbf{c}_{3},\textbf{c}_{4}$  have a $0$ or a 1, respectively. If at some position $i$, $\textbf{c}$ equals $2$, then necessarily $\left\{ c_{1}^{(i)},c_{2}^{(i)}\right\} =\left\{ c_{3}^{(i)},c_{4}^{(i)}\right\} =\left\{ 0,1\right\}.$
If we denote the binary sum by $\oplus$, the former claim is equivalent to $c_{1}\oplus c_{2}=c_{3}\oplus c_{4}$. This contradicts the assumption that $H$ is a parity check matrix of a linear code of distance at least five. Hence, the columns of $H$ must form a $UD_{2}$ code.
\end{IEEEproof}
From the Gilbert-Varshamov bound, one would conclude that $\frac{n(n^{2}-3n+8)}{6}<2^{n-k}$ where $n-k$ is the length of the code and $n$ is the number of codewords. For $n-k$ and $n$ large enough, one would have $n^{3}\approx6\cdot2^{n-k}$ and $n\approx\sqrt[3]{6}\left(2^{\frac{1}{3}}\right)^{n-k}$. 

From the sphere packing bound, one would conclude that $2^{n}\leqslant2^{k}\left[{\sum_{i=0}^{d-1}\binom{n}{i}}\right]$.
%So $2^{n-k}\leqslant1+n+\frac{n(n-1)}{2}+\frac{n(n-1)(n-2)}{6}+\frac{n(n-1)(n-2)(n-3)}{24}.$
For $n\rightarrow\infty$, this implies that $n\geqslant24^{\frac{1}{4}}\left(2^{\frac{1}{4}}\right)^{n-k}$.

Comparing the above method to the construction in \cite{KS64}, one can see that the size of symmetric codes is twice as
large as the size of asymmetric codes.

%%%%%%%%%%%%%%%%%%%%%%%%%%%%%%%%%%%%%%%%%%%%%%%%%%

\end{document}